*******************************************************************
\documentstyle [12pt] {article}
\newcommand{\bfor}{\begin{equation}}
\newcommand{\efor}{\end{equation}}
\newcommand{\ve}{\bf}
\begin{document}
\title{ROLE OF THE LOW-LYING ISOSCALAR DIPOLE MODES
IN THE POLARIZATION POTENTIAL}
\author{ E. B. Balbutsev, A.V. Unzhakova \\
Joint Institute of Nuclear Research, Dubna, Russia\\
\\
   M. V. Andr\'es$^a$, F. Catara$^{b,c}$
    and E. G. Lanza$^{c,b}$\\
             \\
  {$a)$} Departamento de F\'{\i}sica At\'omica y Nuclear,\\
    Universidad de Sevilla, 41080 Sevilla, Spain  \\
  {$b)$} Dipartimento di Fisica dell'Universit\`a,
     95129 Catania, Italy \\
  {$c)$} Istituto Nazionale di Fisica Nucleare,
     Sez. di Catania, \\ 95129 Catania, Italy }
\maketitle
\begin{abstract}
An analysis of the real and imaginary part of the polarization potential in
terms of the relative contributions of the single collective states for the
$ ^{208}Pb + ^{208}Pb $ system has been done. The polarization potential
has been calculated within the Feshbach formalism taking into account the
collective states calculated with the Wigner function moments method. The
contribution of the isoscalar giant dipole resonance states has been
estimated being of the order of 10 - 20 \% of the total at relatively low
incident energy.
\end{abstract}
\newpage
\section { INTRODUCTION }

In the collisions between complex nuclei, the excitation of the internal
degrees of freedom gives rise to a modification of the bare nucleus-nucleus
potential, calculated for instance by the double folding
procedure~\cite{sat}.  This modification is usually referred to as
polarization potential. Many authors~\cite{mau,acl1,acl2,pol}, using
different approaches and various  prescription to calculate the form
factors, have studied the relevance of the excitation of vibrational
collective states (both low lying and giant resonances) for the
determination of the polarization potential. The role played by the
nucleon transfer process has also been investigated~\cite{pol}.

Very recently, by using $(\alpha,\alpha^\prime \gamma)$ reactions, some
experimental evidence has been found~\cite{hol} on the existence  of low
lying isoscalar dipole modes in $^{208}Pb$, $^{90}Zr$, $^{58}Ni$ and
$^{40}Ca$. These modes are of compressional nature and are quite well
described by the method of the Wigner Function Moments (WFM)~\cite{dur}.
Due to  their isoscalar nature, it is interesting to investigate how much
they contribute to the polarization potential. In order to calculate the
latter within the same microscopic approach used in previous
studies~\cite{acl1,acl2} one needs the transition densities of the
considered excited states. As shown in the next section, these can be
extracted from the WFM results by using the linear response theory.
Calculations performed for the  $ ^{208}Pb + ^{208}Pb $ system at various
incident energies show that the contribution of these isoscalar
compressional modes to the polarization potential amounts to 10-20 \%.

\section {  DESCRIPTION OF WFM METHOD }

The starting point for this method is the time-dependent Hartree-Fock
equation for the density matrix $\hat\rho_q({\ve r}_1,{\ve r}_2,t)$:
\begin{equation}
i\hbar{{\partial \hat\rho_q}\over{\partial t}}=[\hat H^{(q)},\hat\rho_q]
\end{equation}
where $\hat H^{(q)}$ is a self-consistent single-particle Hamiltonian;
$q$ is  the isospin index for protons ($q=p$) or  neutrons ($q=n$).
A Skyrme-type effective force (SGII) is used as nucleon-nucleon
  interaction.

Equation (1) is transformed to an equation for the Wigner function
$$f_q({\bf r,p},t)=\frac{1}{2\pi\hbar}\int~ e^{-i{\bf p\cdot s}/\hbar}
{}~\rho_q ({\bf r+s}/2,{\bf r-s}/2,t)~d{\bf s}\,:$$
\begin{equation}
\frac{ \partial f_q}{\partial t}=
\frac{2}{\hbar} \sin{~[\frac{\hbar}{2}~
({\bf\nabla}^H_r.
 {\bf\nabla}^f_p -
 {\bf\nabla}^H_p.
 {\bf\nabla}^f_r)]}
{}~H_W ^{(q)} f_q \quad ,
\end{equation}
where $ H_W^{(q)}({\bf r,p})=\int~ e^{-i{\bf p\cdot s}/\hbar}
({\bf r+s}/2\mid \hat H^{(q)} \mid {\bf r-s}/2)~d{\bf s}$ and
 ${\bf r}=({\ve r}_1+{\ve r}_2)/2,~~{\ve s}={\ve r}_1-{\ve r}_2$.
 The moments of $f_q$ in momentum space define the
nucleon densities $n_q({\bf r},t)$, the mean velocities
 $u_q({\bf r},t)$ and tensors of
 any rank $P_{q\,ij\ldots k}({\ve r},t)$ :
\begin{eqnarray}
n_q({\bf r},t)&=&\int~f_q({\bf r,p},t)~d{\bf p},\nonumber
\\
m\cdot n_q \cdot {\bf u}_q({\bf r},t)&=&  \int~
{\bf p}~ f_q({\bf r,p},t)~d{\bf p},             \nonumber
\\
P_{q\,i_1\ldots i_n}&=& m^{1-n}~\int~(p_{i_1}-m u_{q\,i_1})\ldots
(p_{i_n}-m u_{q\,i_n}) f_q({\bf r,p},t)~d{\bf p}  \nonumber
\end{eqnarray}
where $m$ is the nucleon mass.
 The method consists in taking moments
of equation (2) in phase space, i.e.~in integrating Eq.(2) in ${\bf p}$
and ${\bf r}$ spaces with different weights, in order to obtain a set of
 dynamic equations (virial theorems) for some chosen moments of the
Wigner function.
For example, to describe the negative parity states $1^-,2^-,3^-$ one must
integrate Eq.(2) in momentum space with the weights $1,p_i,p_ip_j,p_ip_jp_k$
($p_i$ are components of the momentum ${\bf p}$) as a first step.
One arrives at a set of coupled time evolution equations for the
density (continuity equation), for the mean velocity $u_{q\,i}({\ve r},t)$
(Euler equation), for the pressure tensor
$P_{q\,ij}({\ve r},t)$ and for the third rank tensor $P_{q\,ijk}({\ve r},t)$.
Then the obtained equations, weighted with $x_i$ and $x_ix_jx_k$ for
$n_q$, with 1 and $x_jx_k$ for $u_{q\,i}$, with $x_q$ for $P_{q\,ij}$
and lastly with 1 for $P_{q\,ijk}$, are integrated in ${\bf r}$-space.
Thus one arrives at a closed system of coupled dynamical equations for
different integral characteristics of the nucleus --
Cartesian tensors of the first and third ranks:
$ \int~n_q({\ve r},t)~x_i~ d{\bf r} $,
$ \int~u_{q\,i}({\ve r},t)~ d{\bf r} $,
$ \int~n_q({\ve r},t)~x_i x_j x_k~ d{\bf r} $,
$ \int~u_{q\,i}({\ve r},t)~x_j x_k d{\bf r} $,
$ \int~P_{q\,ij}({\ve r},t)~ x_k d{\bf r} $,
$ \int~P_{q\,ijk}({\ve r},t)~  d{\bf r} $.

In the case of simple interaction (a harmonic oscillator with a separable
multipole-multipole residual interaction, for example) the integrals
containing the interaction can be expressed in terms of the Cartesian
tensors mentioned above. As a result one obtains a system of
nonlinear equations for the integral characteristics of different
multipolarities.
In the general case of a realistic interaction the integrals containing the
interaction can not be expressed in terms of integral characteristics
without any approximations.
This problem is solved rather easily in the case of small amplitude
motion.
To obtain the corresponding equations we vary the virial equations
taking $ n_q({\ve r},t)=n_q^{(0)}({\ve r})+\delta n_q({\ve r},t) $,
$ u_{q\,i}({\ve r},t)=u^{(0)}_{q\,i}({\ve r})+\delta u_{q\,i}({\ve r},t) $,
$ P_{q\,ij}({\ve r},t)=P^{(0)}_{q\,ij}({\ve r})+\delta P_{q\,ij}
({\ve r},t) $,
$ P_{q\,ijk}({\ve r},t)=P^{(0)}_{q\,ijk}({\ve r})+\delta P_{q\,ijk}
({\ve r},t) $
and ignoring  terms quadratic in the variations $\delta$.
So we arrive at a set of linearised equations, which are a very
convenient means to study the collective small amplitude motion.
These equations are simple differential equations.
In the case of harmonic oscillations they become algebraic equations.
The coefficients in these equations depend only on the ground state
properties. They are linear combinations of integrals over the nuclear
volume of different powers of  ground state particle densities
$n_q^{(0)}$, kinetic energy densities $P_{q\,ii}^{(0)}$ and of their
space derivatives.

\section {  TRANSITION DENSITY }
\label{td}

The transition density is defined as the matrix element
$\langle0|\hat\rho({\ve r})|\alpha\rangle$ of the density operator
\bfor
\hat\rho({\ve r}) =  \sum_{i=1}^N \delta ({\ve r}-\hat r_i)
\efor
where $\hat r_i$ is the position operator of particle $i$ and $|0\rangle$ and
$|\alpha\rangle$ are the ground and excited state, respectively~\cite{r&s}.
 The problem of WFM method consists in finding this matrix element without
calculating the wavefunctions of $|0\rangle$ and $|\alpha\rangle$. This
 can be solved
with the help of linear response theory. The linear response to the
perturbation operator $\hat O (t) = \hat F \exp {(-i\omega t)} + \hat
F^\dagger \exp {(+i\omega t)}$ can be written as
\begin{eqnarray}
\rho^{(1)}_{kl} &=& {1 \over \hbar} \sum_{pq\nu}
        \Big\{ {\langle0|a^\dagger_l a_k|\nu\rangle
 \langle\nu|a^\dagger_p a_q|0\rangle
	\over {\omega - \Omega_\nu} } -
        {\langle0|a^\dagger_p a_q|\nu\rangle
\langle\nu|a^\dagger_l a_k|0\rangle
	\over {\omega + \Omega_\nu} } \Big\} f_{pq} \nonumber \\
       &=& {1 \over \hbar}
        \sum_\nu \Big\{ {\langle0|a^\dagger_l a_k|\nu\rangle
\langle\nu|\hat F|0\rangle
	\over {\omega - \Omega_\nu} } -
        {\langle0|\hat F|\nu\rangle \langle\nu|a^\dagger_l a_k|0\rangle
	\over {\omega + \Omega_\nu} } \Big\}
\end{eqnarray}
where $a^\dagger_l, a_k$ are creation and annihilation operators,
$\Omega_\nu$ the eigenfrequencies of the system, and $\hat F = \sum _{pq}
f_{pq} a^\dagger_p a_q$. Using the second quantization representation for
the density operator
\bfor
\hat \rho ({\ve r}) = \sum_{qp} d_{pq} ({\ve r}) a^\dagger_p a_q \, ,
\efor
where $d_{pq} ({\ve r}) = \langle p| \delta ({\ve r}-\hat r)|q\rangle =
 \phi_p^\ast ({\ve r}) \phi_q
({\ve r})$ and $\phi_q ({\ve r})$ being a basis of single particle
wavefunctions we can write the following equation for the change of density
\begin{eqnarray}
\delta n ({\ve r}) &=& \sum_{kl} \rho_{kl}^{(1)} d_{lk} ({\ve r})
 \nonumber \\
                  &=&{1 \over \hbar}
  \sum_\nu \Big\{ {\langle 0|\hat\rho ({\ve r})|\nu\rangle
\langle\nu|\hat F|0\rangle
  \over {\omega - \Omega_\nu} } -
 {\langle 0|\hat F|\nu\rangle \langle\nu|\hat\rho ({\ve r})|0\rangle
  \over {\omega + \Omega_\nu} } \Big\} \,, \label{eq:dn}
\end{eqnarray}
which we calculate with the help of the WFM. By varying the continuity
equation
\begin{equation}
\frac{\partial n_q}{\partial t}=-div\,\{n_q {\ve u}_q+\eta n_q n_{q'}
({\ve u}_q-{\ve u}_q')\}
\end{equation}
 we express the change of density in  terms of small displacements
$\xi_{q\,i}$
\bfor
\delta n_q = - \sum_{s=1}^3\, {\partial\over{\partial x_s}}\,
 \{ n_q  \xi_{q\,s} + \eta \, n_q n_{q^\prime}
             ( \xi_{q\,s} -  \xi_{q^\prime\,s}) \}
        \label{eq:dn2}
\efor
where ${ \partial \over {\partial t}} \xi_{q\,i} ({\ve r},t) =
\delta u_{q\,i} ({\ve r},t)$.
These displacements are parametrized in the following way
\bfor
\xi_{q\,i} ({\ve r},t) = L_{q\,i}(t) + \sum_{j=1}^3 L_{q\,i,j}(t)\, x_j +
                     \sum_{j,k=1}^3 L_{q\,i,jk}(t)\, x_j x_k
\efor
The tensors $L_{q\,i}(t)$, $L_{q\,i,j}(t)$, $L_{q\,i,jk}(t)$ are found by
 solving
the system of coupled dynamical equations involving moments of the Wigner
function. Their derivation was described in ref.~\cite{dur,Piper}. It is
evident that one can extract the information about the transition density from
eq. (\ref{eq:dn}) taking the limit
\bfor
\lim_{\omega \rightarrow \Omega_\alpha} \hbar (\omega - \Omega_\alpha)
    \delta n({\ve r}) =
\langle 0|\hat\rho ({\ve r})|\alpha\rangle \langle\alpha|\hat F|0\rangle\,.
        \label{eq:lim}
\efor
Thus, to determine the transition density
 $\langle0|\hat\rho({\ve r})|\alpha\rangle$ from
this expression it is necessary to know the matrix element
 $\langle\alpha|\hat F|0\rangle$. By using eqs. (\ref{eq:dn}) and
(\ref{eq:lim})
 one can get the square modulus of the required matrix element :
\bfor
\lim_{\omega \rightarrow \Omega_\alpha} \hbar (\omega - \Omega_\alpha)
        \int \, d{\ve r} \, f({\ve r}) \delta  n({\ve r}) =
        \langle0| {\hat F} |\alpha\rangle \langle\alpha| {\hat F} |0\rangle=
        \big| \langle\alpha | \hat F |0\rangle \big| ^2 \, .
\efor
which is true for any hermitian single particle operator $\hat F$. In the
case of a real operator one can write
\bfor
\langle\alpha| \hat F |0\rangle = \pm \Big[
\lim_{\omega \rightarrow \Omega_\alpha}
	\hbar (\omega - \Omega_\alpha)
        \int \, d{\ve r} \, f({\ve r}) \delta n ({\ve r})
 \Big]^{1\over 2} \, .
\efor
Then the final expression for the transition density is
\bfor
\langle0| \hat \rho ({\ve r}) |\alpha\rangle =
	{\pm  \lim_{\omega \rightarrow \Omega_\alpha}
        \hbar (\omega - \Omega_\alpha) \delta n({\ve r}) \over
	\big[ \lim_{\omega \rightarrow \Omega_\alpha}
	\hbar (\omega - \Omega_\alpha)
        \int \, d{\ve r} \, f({\ve r}) \delta n ({\ve r})
\big]^{1\over 2} }\, .
	\label{eq:fdn}
\efor
Combining eqs. (\ref{eq:fdn}) and (\ref{eq:dn2}) we get the explicit
expressions
for the transition densities of the various multipolarities
\begin {eqnarray}
\langle0| \hat \rho ({\ve r}) |2^+\rangle &=& r {\partial n(r)
 \over \partial r}  (A_1 + A_2 n(r)) Y_{2\mu} \, ,
\label{eq:r1} \\
\langle0| \hat \rho ({\ve r}) |3^-\rangle &=& r^2 {\partial n(r)
 \over \partial r}   (B_1 + B_2 n(r)) Y_{3\mu} \, ,
\label{eq:r2} \\
\langle0| \hat \rho ({\ve r}) |1^-\rangle &=& \Big\{ r \,
 n(r) (C_1 + C_4 n(r)) + \nonumber \\
			& & {\partial n(r) \over \partial r}
			    [C_3 + C_2 \, r^2 + n(r) (C_5 + C_6 r^2)]
			    \Big\} Y_{1\mu} \, , \\
\langle0| \hat \rho ({\ve r}) |0^+\rangle &=& \Big\{ D_1 \big( n(r) +
 {r \over 3}    {\partial n(r) \over \partial r} \big) +
			    \nonumber \\
			& & D_2 \big( n^2 (r) + {2 \over 3} r \, n(r)
			    {\partial n(r) \over \partial r} \big)
			    \Big\} Y_{00}
\end {eqnarray}
where the constants
$A$ and $D$ are combinations of the tensors $L_{q\,i,j}$,
$B$ - of tensors $L_{q\,i,jk}$, $C$ - of tensors $L_{q\,i}$ and
 $L_{q\,i,jk}$.
These expressions are too complicate, to be written here.
So we describe  the derivation of formulae for
the most simple constants ($A_1$, $A_2$) in the Appendix,
just to give a general idea of the procedure.

The transition density of Giant Octupole Resonance (GOR) differs from that
of Low Energy Octupole Resonance (LEOR) by the values of the  constants
$B$. It is necessary to note that in our calculations LEOR includes also
the contribution from the lowest $3^-$ state, because the WFM method gives
the centroid of all $3^-$ states lying below GOR. The expressions
(\ref{eq:r1}), (\ref{eq:r2}) for the transition  densities of $2^+$ and
$3^-$ excitations differ from that of Tassie model by terms quadratic in
the density (terms in $A_2$ and $B_2$). Numerically these terms turn out to
be small. The formula for the transition density of $0^+$ practically
coincides with the one derived in~\cite{VanG}. The only difference is the
term proportional to $D_2$, which however is numerically very small. The
expression (16) can be compared with formula of Harakeh and Dieperink
{}~\cite{C},  which was derived supposing that all energy weighted sum rule
(EWSR) is exhausted by one state. All the three terms linear in the density
have the same radial dependence, but the corresponding coefficients
$C_1,~C_2,~C_3$ have nothing in common with that of ref.~\cite{C}.
Moreover, the terms proportional to $n(r)^2$ give a noticeable contribution
in this case. So the total radial dependence of $1^-$ transition density
may be different from that of ~\cite{C}. We have three low-lying $1^-$
states contributing to the isoscalar EWSR. Their proton and neutron
transition densities are shown in Fig.1, left and right side, respectively.
The dashed lines give the transition densities calculated by using only the
linear terms in the density, while the solid lines include also the
quadratic ones. It is seen that the transition densities of all three
excitations are rather large inside the nucleus demonstrating their
compressional nature. The highest excitation has the largest contribution
quadratic in the density and has a rather big admixture of isovector mode.

\section { POLARIZATION POTENTIAL }
\label{popo}

The most elegant way to calculate the polarization potential is to follow
the Feshbach formalism~\cite {mau,acl2}. In this approach is very easy to
take explicitly  into account the contribution of some definite collective
states. Here we will briefly describe the method which has been extensively
described in ref.~\cite {mau,acl2}. In the Feshbach approach the effective
heavy ion  interaction is written as
\begin {eqnarray}
V ({\ve R},{\ve R}^\prime)& =&\langle00| v({\ve R})|00\rangle
   \delta ({\ve R}-{\ve R}^\prime)         \nonumber \\
&+&
   \sum_{K_1 K_2} \langle00|v({\ve R})|K_1 K_2\rangle 
 \times\, G_{K_1 K_2}({\ve R}, {\ve R}^\prime)
   \langle K_1 K_2|v({\ve R}^\prime)|00\rangle  \nonumber \\
 & = & V_F ({\ve R}) + \Delta {\cal V} ({\ve R},{\ve R}^\prime) \quad .
	\label{eq:pp}
\end {eqnarray}
In the first term the nucleon-nucleon interaction $v({\ve R})$ is double
folded with the ground state densities of the two nuclei. In the second
term the sum is over all the non-elastic channels,  and  $v ({\ve R})$ is
double folded with the transition densities of the two nuclei:  it
describes the coupling of the elastic channel to the non-elastic ones. This
term is the so called dynamical polarization potential. Its physical
meaning is transparent from  eq. (\ref{eq:pp}): The interaction acting at the
distance ${\ve R}^\prime$ takes the system in one of the eliminated
channels, then it is propagated at another distance ${\ve R}$ where the
interaction, acting again, brings back the system into the elastic channel.
Then the polarization potential is non-local, and if one or more of the
eliminated channels is open it is also complex and its absorptive part
describes the loss of flux from the elastic channel. The couplings make
$\Delta {\cal V}$ also energy dependent because of the appearance of the
energy in the propagator $G_{K_1 K_2}({\ve R}, {\ve R}^\prime)$.

The calculation of eq. (\ref{eq:pp}) is a very difficult task because of the
presence of the propagator, then we calculate it in the WKB approximation:
\bfor
G_{K_1 K_2}(\rho,s) \simeq -{\mu \over 2 \pi \hbar^2}
    {exp (i M_{K_1 K_2} (\rho) s) \over s}
\efor
with
\bfor
M^2_{K_1 K_2} (\rho) = {2 \mu \over \hbar^2}\,
  [E_{cm}-E_{K_1}-E_{K_2}-V_L(\rho)-V_C(\rho)]
\efor
where
\bfor
{\ve \rho}= {1 \over 2}({\ve R} + {\ve R}^\prime)  \, ; \quad
  {\ve s}= |{\ve R} - {\ve R}^\prime|
\efor
The local optical potential $V_L(\rho)$ is given by
\bfor
V_L(\rho) = V_F(\rho) + \Delta {\cal V}_L (\rho)
\efor
while $V_C$ is the Coulomb potential between the two nuclei. A procedure to
define the local polarization potential $\Delta {\cal V}_L$ will be given
later on in this paper.

Within the double folding approach the form factors can be written as
\bfor
F_{K_1 0}({\ve R})\equiv \langle00|v({\ve R})|K_10\rangle =
                  \int \,d{{\ve r}_1} d{{\ve r}_2}\,
                  \rho_{_{K_10}}({\ve r}_1) v(|{\ve r}_1- {\ve r}_2+
                        {\ve R}|)
                  \rho_{_{00}}({\ve r}_2)
	\label{eq:ff}
\efor
where $\rho_{_{K_10}}$ and $\rho_{_{00}}$ are the transition density and
the ground state density, respectively. These are calculated within the WFM
method, as described in section~\ref{td}. The details of the calculation of
$\Delta {\cal V}({\ve \rho}, {\ve s})$ is reported in ref.~\cite{acl2}.

The potential so obtained is non-local, but since the range of non-locality
of $\Delta {\cal V}$ is small with respect to its radius~\cite{acl2}
we can  use a standard procedure~\cite{per} to obtain a local potential
from a non-local one. Then
\begin {eqnarray}
\Delta V (\rho) & = & 4 \pi \int j_{_0} (k s) Re \Delta {\cal V}(\rho,s)
                    s^2 ds \\
W (\rho) & = & 4 \pi \int j_{_0} (k s) Im \Delta {\cal V}(\rho,s)
                    s^2 ds
\end {eqnarray}
where $j_{_o}$ is the Bessel function and
\bfor
k = {2\mu \over \hbar^2}\, [E_{cm}-V_F(\rho)-V_C(\rho)]\quad . \nonumber
\efor

\section {RESULTS AND DISCUSSION}

We have done calculations for the system $ ^{208}Pb + ^{208}Pb $ at several
incident energies.   The levels used in the calculations are reported in
Table I : they were obtained with the WFM method. We have
parametrized the ground state density with a Fermi distribution  $n(r) =
n_0/(1 + \exp {[(r-R)/a])}$ whose parameters are the following
{}~\cite{Bern}: $R =1.115\, A^{1/3}-0.53\,A^{-1/3}~fm $, $a = 0.568~fm$,
while  $n_0 $ is fixed by the condition $ 4\pi\int^\infty_0 n(r) r^2\,
dr=A$. As already stated in section~\ref{td}, the transition densities have
the advantage to have an analytical form from which one can see that their
first term corresponds to the Tassie model in the case of $2^+$, $3^-$ and
$0^+$ multipolarities. The corrections with respect to the Tassie ones are
not important. Formula for $1^-$ is new and has nothing common with Tassie
model, because it takes into account the center of mass movement.
The bare potential has been obtained by double folding the nucleon -
nucleon effective interaction M3Y~\cite{sat} with the ground state density
of the two nuclei. In the same way we have obtained the form factors eq.
(\ref{eq:ff})  by double folding the M3Y with the ground state density of
one nucleus and the transition density of the other one. Examples of form
factors are shown in Fig. 2: on the right side we show the one related to
the LEOR, while on the left side are reported the ones concerning the three
isoscalar GDR states. In the latter case the oscillations are due to  the
fact that the form factors change sign around $10 fm$.

In order to show the relative contribution of the states included in the
calculations we have computed the polarization potential for various
energies as function of the relative distance $R$. In fig. 3 we show the
contribution due to different multipolarities( as indicated in the figure)
to the real (left) and imaginary (right) part of the polarization potential
plotted in percentage of the total potential as function of the relative
distance $R$. We notice that the most important contribution to both real
and imaginary part is due to the low lying $3^-$ state. As the incident
energy is increasing the relative contribution of the GQR becomes more
important. This different behaviour has already been found in previous
studies~\cite{acl1,acl2}. The novel result is represented by the
contribution of the isoscalar GDR states which lies between 10-20 \%, that
is to say comparable to the contribution of the GQR at least for the
imaginary.

We have also calculated the polarization potential at a fixed relative
distance ($R~=~16~fm$) as a function of the incident energy taking into
account the explicit contribution of the different multipolarities. The
results are shown in fig. 4 where the different curves correspond to the
contribution of the different states as indicated in the figure. Again we
notice the importance of the LEOR, which is overwhelming with respect to
the others states in a large range of incident energy for the case of the
imaginary part, while for the real part the contribution of the GQR, at
higher energies, is essentially the same than the one of the low lying
$3^-$ state. The contribution of the isoscalar GDR to both real and
imaginary part of the polarization potential is more important (at this
distance) at low energy, and it decreases with the energy slower for the
absorbitive part than for the real part.

The interplay between the LEOR and the GQR states seems to contradict the
findings of ref.~\cite{acl1,acl2}. In particular the contribution of the
LEOR  to the absorbitive part is very big with respect to the one of the
GQR. As explicitly stated in the second paper of ref.~\cite{acl2}, this is
due to the fact that we are using different transition densities (hence
different form factor) from the one used in ref.~\cite{acl2} which were
calculated in the framework of the Random Phase Approximation.

\section { CONCLUSION }
We have done an analysis of the real and imaginary part of the polarization
potential in terms of the relative contributions of the single collective
states for the system $ ^{208}Pb + ^{208}Pb $. The polarization potential
has been calculated within the Feshbach formalism taking into account the
collective states calculated with the WFM method. Within this method it is
possible to obtain also isoscalar GDR states whose contribution to the
polarization potential has been estimated being of the order of 10-20 \% of
the total at relatively low incident energy. Then this  contribution should
in principle show up in some experimental observable, like for instance the
elastic cross section. The other multipolarities give a contribution which
has been analyzed in great details in previous work~\cite{acl1,acl2} where
slightly different results were obtained. This can be  ascribed to the
different form factor we have used in the two cases.

\newpage
\begin{center}
APPENDIX
\vspace{0.8mm}

Transition density for $2^+$ excitation.
\end{center}

Multiplying (8) by $r^2Y_{2\mu}(\theta,\varphi)$ and integrating over angles
we find the radial dependence for the change of density
$$
\delta n_q(r)_{2\mu}=-\frac{4\pi}{15}\{r\frac{\partial n_q}{\partial r}
\cdot {\cal L}_{q\,2\mu}+\eta r \frac{\partial}{\partial r}
(n_q\cdot n_{q^\prime})\cdot({\cal L}_{q\,2\mu}-
{\cal L}_{q^\prime\,2\mu})\}\,,
\eqno(A1)
$$
where the components of the irreducible tensors ${\cal L}_{q\,2\mu}$ are linear
 combinations of $L_{q\,ij}$. In the case of a spherical nucleus we can
study any component of ${\cal L}_{q\,2\mu}$, for example  ${\cal L}_{q\,22}$:
$$
{\cal L}_{q\,22}=L_{q\,11}-L_{q\,22}+2iL_{q\,12},~~~~L_{ij}=L_{i,j}+L_{j,i}.
\eqno(A2)
$$
For a spherical nucleus $L_{q\,11}=L_{q\,22}$ so we will concentrate our
attention on $L_{q\,12}$. Following the prescriptions of chapter 2 and  using
formulae (8,9) one can derive the system of coupled dynamical equations for
$L_{q\,12}$ and $\int \delta P_{q\,12} d{\ve r}\equiv \pi_{q\,12} $
 (see ref.~\cite{Piper}):
$$
\ddot{L}_{q\,12}-2b_{q\,1}L_{q\,12}+2b_{q\,2}L_{q^\prime\,12}-2b_{q\,3}
\beta\pi_{q\,12}=
$$
$$
-i\beta\int r \frac{\partial n_q}{\partial r}
\,Y_{2\mu}\cdot W(t)\,d{\ve r}\,,
\eqno(A3)
$$
$$
\beta\dot\pi_{q\,12}
+d_{q\,1}\cdot \dot{L}_{q\,12}-
d_{q\,2}\cdot \dot{L}_{q^\prime\,12}=0\,,
$$ where
$ \beta=\frac{3A}{4\pi m Z_q \alpha_1^4}\ $
and $W(t)=e_p r^2
(Y_{2\mu}e^{-i\omega t}+Y_{2\mu}^*e^{i\omega t}) $
 is the external field.The corresponding system for the other kind of
nucleons is obtained by changing
$q\leftrightarrow q'$.  The following notations have been
introduced here:
$$
b_{q\,1}=\frac{z'}{m\alpha^4_1}{\cal A}+
\gamma\frac{t_+}{4\hbar^2\alpha_1^4}{\cal E}_{q^\prime}+
2T_{3q}+z'(7T_1+2T_{2q})+\delta_{q,p}\cdot \frac{8}{15}
\varphi(1+\eta n_0 z')\,,
$$
$$
b_{q\,2}=\frac{z}{m\alpha^4_1}{\cal A}+
\gamma\frac{t_+}{4\hbar^2\alpha_1^4}{\cal E}_{q}
+z(7T_1+2T_{2q})+\delta_{q,p}\cdot \frac{8}{15}\varphi\eta n_0 z\,,
$$
$$
b_{q\,3}=1+\frac{2\pi m}{\hbar^2 A}\alpha_2^2\cdot (t_+-z t_-/2),
{}~~d_{q\,2}=\frac{15}{8}\gamma\eta z \frac{\alpha^2_{8/3}}{m\alpha_1^4}\,,
$$
$$
d_{q\,1}=\frac{3\gamma}{m\alpha_1^4}(\alpha_{5/3}^2+\frac{5}{8}\eta z'
\alpha^2_{8/3})\,,
$$
\begin{eqnarray}
{\cal A}&=&\frac{t_0}{5}(1+\frac{x_0}{2})(S_2^4+2\eta{\cal R}^1_4)+
\frac{t_3}{120}\{(\sigma+1)[(\sigma(1-x_3)+2(2+x_3)]+ \nonumber  \\
&&+2\sigma(\sigma-1)(1+2x_3)z z'\}\cdot({\cal R}_4^{\sigma}+2\eta
{\cal R}_4^{\sigma+1})\,,
\nonumber
\end{eqnarray}
$$
{\cal E}_q=z({\cal R}_4^{2/3}+2\eta{\cal R}_4^{5/3})\,,
{}~~\varphi=\pi n_0\frac{e_p^2}{m}\frac{Z_p}{A}\,,
$$
$$
T_1=\frac{\hbar^2}{4m\alpha_1^4}\frac{t_-}{35}[4S_2^2+Q_0+2\eta(4{\cal R}_2^1
+Q_1-\frac{4}{3}S^3_3)]\,,
$$
$$
T_{q\,2}=\frac{\hbar^2}{4m\alpha_1^4}\eta
(\frac{z}{2}t_+-t_-)
({\cal R}_2^1
+\frac{1}{5}S^3_3)]\,,~~
T_{q\,3}=\frac{\hbar^2}{8m\alpha_1^4}
(\frac{z}{2}t_+-t_-)
S^2_2\,,
$$

$$
\alpha_\mu^\nu=\int_0^\infty n(r)^\mu r^\nu\,dr,~~~
{\cal R}_\nu^\mu=\int_0^\infty n(r)^\mu r^\nu
 (\frac{\partial n}{\partial r})^2\,dr
$$
$$
S^\mu_\nu=\int_0^\infty r^\mu
 (\frac{\partial n}{\partial r})^\nu\,dr,
{}~~Q_\mu=\int_0^\infty n(r)^\mu r^4
 (\frac{\partial^2 n}{\partial ^2r})^2\,dr\,,
$$
$$
n(r)=n_p(r)+n_n(r),~~~z=\frac{Z_q}{A},~~~z'=\frac{Z_{q'}}{A},
{}~~~\eta=\frac{mt_+}{2\hbar^2},~~~t_+=t_1+t_2,
$$
$
t_-=t_1-t_2$. The quantities
$t_0,~t_1,~t_2,~t_3,~\sigma,~x_0,~x_3$ are the
 parameters of Skyrme forces, $e_p$ the proton charge, $m$ the  nucleon mass.
Deriving these formulae for $b_{q\,i},\ d_{q\,i}$ we used the Thomas-Fermi
approximation for the ground state pressure tensor
 $P^{0}_{q\,ij}({\ve r})=\delta_{ij} \gamma n_q(r)^{5/3} $ with
 $\gamma=\frac{\hbar^2}{5m}(3\pi^2)^{2/3} $.

The system of differential equations (A3) become algebraic when
one takes into account the evident time dependence of variables
 $ L_{q\,12}(t)=\tilde{L}_{q\,12}\,e^{i\omega t}$
and $\pi_{q\,12}(t)=\tilde\pi_{q\,12}\, e^{i\omega t} $.
 the solution of this system is :
$$\tilde{L}_{q\,12}=\Delta_q/\Delta,~~\tilde{L}_{q'\,12}=\Delta_{q'}/
\Delta\,,
\eqno(A4)
$$
where $\Delta$  is the determinant of system (A3). The determinant
$\Delta_q$ is obtained by substituting the corresponding column of
$\Delta$ by the right hand side column of (A3).
 $\Delta$ is a biquadratic polynomial and has two roots
 $\Omega_s^2$ and $\Omega_v^2$ which are
 interpreted as isoscalar and isovector giant quadrupole resonances.
Writing the determinant in the form
 $\Delta=c\cdot(\omega^2-\Omega^2_s)(\omega^2-\Omega^2_v) $,
 where $c$ is a constant
and putting (A4) into (A1) one can  calculate easily the limits
in (13) and find the constants $A_1,~A_2$.

\newpage
\begin{center}
FIGURE CAPTION
\end{center}

	Fig.1\\
Proton (left) and neutron (right) transition densities for low-lying dipole
excitations in $^{208}Pb$ with the energies $E_1=7.65 MeV$, $E_2=8.99 MeV$
and $E_3=10.01 MeV$, from top to bottom.  The dashed curves are the results
of calculations with only the linear density terms (formulae (14)-(17)).\\

	Fig.2\\
Form factors for $^{208}Pb$ for the reaction $^{208}Pb + \,^{208}Pb$ for
two different multipolarities. The three curves in the left part correspond
to the isoscalar GDR states with excitations energies $E_1=7.65~MeV$ (solid
curve), $E_2=8.99~MeV$ (dashed curve) and $E_3=10.01~MeV$ (long - short
dashed curve).\\

	Fig.3\\
Contribution of the different multipolarities ( as indicated in the figure)
to the real (left) and imaginary (right) part of the local polarization
potential plotted in percentage of the total potential as function of the
relative distance $R$ for three different values of incident energies for
the reaction $^{208}Pb + \,^{208}Pb$.\\

	Fig.4\\
Contribution of the different multipolarities ( as indicated in the figure)
to the real (upper part) and imaginary (lower part) part of the local
polarization potential plotted as function of incident energy for a fixed
relative distance $R = 16 fm$ for the reaction $^{208}Pb + \,^{208}Pb$.\\

\newpage

	\begin{center}
	Table 1. \\
   Isoscalar collective states of $^{208}Pb$ used in the calculations.\\
	\end{center}
	\begin{center}
 	\begin{tabular}{||c|r|r||}            \hline
  	\multicolumn{1}{||c|}{$J^\pi$ } &
  	\multicolumn{1}{|c|}{$E,\,MeV$} &
  	\multicolumn{1}{|c||}{$EWSR,\,\%$} \\  \hline
  	$1^-$  & 7.65   & 5.79     \\
        	& 8.99   & 2.94    \\
         	& 10.01  & 1.13    \\   \hline
  	$2^+$  & 11.82  & 33.35    \\   \hline
  	$3^-$  & 4.65   & 15.14    \\
         	& 22.21  & 16.20   \\   \hline
  	$0^+$  & 13.63  & 34.28
  	\\[1.5mm]
 	\hline
 	\end{tabular}
	\end{center}

\end{document}